\documentclass{aa} 

\usepackage{graphicx}  
\usepackage{txfonts}  
\usepackage{natbib}
\bibpunct{(}{)}{;}{a}{}{,}
\usepackage{xcolor}
\usepackage{soul} 

\begin{document}  
\newcommand{\felipe}[1]{\bf #1}
\newcommand{\sylvie}[1]{\bf #1}
\newcommand{\gerard}[1]{\bf #1}
\newcommand{\sout}[1]{\textst{#1}} 
\title{New simulations of accreting DA white dwarfs: inferring 
accretion rates from the surface contamination}
  
  \author{ F. C. Wachlin\inst{1},
           G. Vauclair\inst{2,3},
           S. Vauclair\inst{2,3},
           \and
           L. G. Althaus\inst{1}
           }
\institute{Instituto de Astrof\'{\i}sica de La Plata (UNLP - CONICET). 
           Facultad de Ciencias Astron\'omicas y Geof\'{\i}sicas. 
           Universidad Nacional de La Plata, Argentina\\
           \email{fcw@fcaglp.unlp.edu.ar}     
           \and
           Universit\'e de Toulouse, UPS-OMP, IRAP, France
           \and
           CNRS, IRAP, 14 avenue Edouard Belin, 31400 Toulouse, France\\
           }
\date{\today}

\abstract{
A non negligible fraction of white dwarf stars show the presence
of heavy elements in their atmospheres. The most accepted explanation
for this contamination is the accretion of material coming from
tidally disrupted planetesimals, which form a debris disk around the 
star.}
{We provide a grid of models for hydrogen rich white dwarfs accreting
heavy material. We sweep a 3D parameter space involving different 
effective temperatures, envelope's hydrogen content and accretion
rates. The grid is appropriate for determining accretion rates   
in white dwarfs showing the presence of heavy elements.}
{Full evolutionary calculations of accreting white dwarfs were computed including all relevant physical processes, 
particularly the fingering (thermohaline) convection, a process
neglected in most previous works, that has to be considered to obtain
realistic estimations. Accretion is treated as a continuous process
and bulk Earth composition is assumed for the accreted
material.}
{We obtain final (stationary or near stationary) and reliable 
abundances for
a grid of models representing hydrogen rich white dwarfs of
different effective temperatures and hydrogen contents, applied
to various accretion rates.}
{Our results provide estimates of accretion rates, accounting
for thermohaline mixing, to
be used for further studies on evolved planetary systems.}

  
\titlerunning{New simulations of accreting DA white dwarfs}
  
\authorrunning{Wachlin et al.}  

\maketitle 

\section{Introduction}  
\label{intro}

All the stars with masses lower than 8 $M_\odot$, which constitute
about 97{\%} of the stellar population of the Galaxy, will end their 
evolution as white dwarfs \citep{1997ApJ...489..772I}.
A large fraction of these stars host planets \citep{2012Natur.481..167C}.
The fate of these planetary systems, when the stars 
evolve from the main-sequence up to the final white dwarf stage, has been the 
subject of considerable interest during the last decades
\citep{Debes2002,Debes2012,Mustill2013,Veras2013,Frewen2014}.

The infrared 
excess discovered around the DA white dwarf G29-38
\citep{Zuckerman1987} and the photospheric contamination 
of white dwarfs by heavy elements are interpreted as the result of 
the disruption by tidal effects of planetesimals orbiting the white 
dwarf \citep{Jura2003}. This scenario is confirmed by the observations 
of debris transiting the white dwarf WD1145+017 \citep{Vanderburg2015}
and the spectroscopic detection of planetesimal orbiting in the 
gaseous disk of SDSS1228+1040 \citep{Manser2019}. These observations
show that small bodies in the planetary systems have survived during the 
host-star evolution. This implies that some planets must have survived as well. 
Such massive bodies are needed to perturb the orbits of the planetesimals 
and push them inside the white dwarf tidal radius where they disintegrate, 
as predicted by most scenarios of planetary systems evolution
\citep{Veras2013,Veras2014a,Veras2014b,Veras2015a,Veras2015b,Veras2015c,
Veras2016a,Veras2016b}. 

The disrupted planetesimals feed the debris disk, and some of it is accreted 
onto the white dwarf, thereby polluting its atmosphere. 
Since the diffusion time-scale of the accreted heavy elements through the 
white dwarf external layers is much shorter than the evolutionary time-scale, 
the presence of heavy elements in the photosphere implies that the accretion 
process is ongoing. The study of polluted white dwarfs is accordingly a 
powerful way to study the chemical composition of the planetesimals and to 
better understand the various physical processes at work in the evolution 
of planetary systems. 

The estimate of the accretion rates is an important input in this study.  
Most previous estimates of accretion rates were obtained by assuming 
that the accreted  material, completely mixed in the surface convection zone, 
diffuses downwards on a diffusion time scale 
\citep{1992ApJS...82..505D, 2009A&A...498..517K,2012MNRAS.424..464F,
2014A&A...566A..34K}. Recently, \cite{2019MNRAS.488.2503C} explored the 
macroscopic diffusion induced by convective overshoot in DA white dwarfs 
by using three-dimensional (3D) radiation hydrodynamic simulations 
with the CO5BOLD code \citep{2012JCoPh.231..919F}. 
They found the mixed mass may increase by up to 2.5 dex, where such 
an increase of the mixed region leads to accretion 
rates which are a factor 2--5 larger.

\cite{2013A&A...557L..12D,2017A&A...601A..13W} and 
\cite{2018ApJ...859L..19B,2019ApJ...872...96B} introduced in their computations 
the fingering convection process, which is indeed unavoidable in this 
context, since the accreted material, with a chemical composition mostly 
similar to that of the solar system bodies \citep{2019MNRAS.490..202S}
has a mean molecular weight larger than that of the white dwarf atmospheres. 
The inverse $\mu$-gradient produces a double-diffusive instability, 
inducing extra-mixing of the accreted material (see for example 
\cite{2004ApJ...605..874V,2007A&A...464L..57S,2011ApJ...728L..30G,
2011A&A...533A.139W,2013ApJ...768...34B,2014A&A...570A..58W,
2014ApJ...795..118Z}).

As shown by \cite{2013A&A...557L..12D} and confirmed by 
\cite{2017A&A...601A..13W} and \cite{2018ApJ...859L..19B,2019ApJ...872...96B}, 
this fingering convection has important consequences in the case of accreting 
DA white dwarfs, whereas it is absent or marginal in DB white dwarfs. 
In DA white dwarfs, the accretion rates needed to reproduce the photospheric 
abundances of heavy elements exceed by up to 2 orders of magnitude those 
estimated without this effect. 

In this paper, we present the results of a series of numerical simulations 
of accretion onto DA white dwarfs. Our aim is to provide estimates of 
the accretion rates and of the photospheric chemical composition for a 
choice of heavy elements among the most often observed in polluted white 
dwarfs. Our simulations cover a large range of parameters for the effective 
temperature, hydrogen mass fraction and accretion rates. 
Due to computation time limitations, we have to restrict 
ourselves to study only one white dwarf's mass. From these results, 
it is possible to infer an estimate of the accretion rate needed to 
reproduce the heavy element abundances deduced from the observations, 
knowing the effective temperature of the white dwarf. In section 2 we 
define the range of parameters covered by the simulations and describe 
how we obtain the initial models. Section 3 describes how the simulations 
have been performed. Section 4 gives the results of our simulations. 
A summary and a discussion of these results are given in section 5.

\section{Initial models}  
\label{sec:modini}

To study the relation between the accretion rate and the resulting
surface contamination of hydrogen-rich (DA) white dwarfs, we prepared
a set of numerical experiments involving models of white dwarfs with
different effective temperatures and different amounts of hydrogen
content in their envelopes, ($M_\mathrm{H}$). In particular, we choose
the following  effective temperatures:  6000K, 8000K, 10000K, 10500K,
11000K, 11500K, 12000K, 16000K, 20000K and 25000K and the following
$M_\mathrm{H}$ values: $10^{-4} M_\odot$, $10^{-6} M_\odot$,
$10^{-8}M_\odot$ and $10^{-10} M_\odot$. The value of $M_\mathrm{H}$
is particularly relevant since it impacts the depth of the transition
zone between hydrogen-rich and helium-rich layers. 

Our parameter space partially overlaps that of 
\cite{2019ApJ...872...96B}, which spans the following ranges:  
$6000 \mathrm{K} < T_\mathrm{eff} < 20000 \mathrm{K}$,
$M_\mathrm{WD}/M_\odot = 0.38, 0.60, 0.90$ and
$\dot M= 10^4$ g s$^{-1}$--$10^{12}$ g s$^{-1}$, 
for a fixed hydrogen content in the envelope of 
$M_\mathrm{H}=10^{-6} M_\mathrm{wd}$.

All initial setups are based on the 0.609 $M_\odot$ ($Z=0.01$)
white dwarf model obtained by \cite{2010ApJ...717..183R} from the full
evolution of its progenitor star from the zero-age main sequence
(ZAMS) to advanced stages on the thermally-pulsing asymptotic giant
branch. To generate white dwarf configurations with smaller hydrogen
contents than that dictated by progenitor evolution, we artificially
reduced the hydrogen content by converting the excess of hydrogen into
helium. This is enough for our purposes. In this work, we considered
for each model four different accretion rates, namely $\log (\dot M)
= 6, 8, 10, 12$, where $\dot M$ is given in g.s$^{-1}$. 

\section{Numerical simulations}  
\label{simulations}

The white dwarf models used in this work were generated by
the {\tt  LPCODE} stellar evolution code. This code has been tested
and widely used in various stellar evolution contexts of low-mass and
white dwarf stars 
\citep[see][for details]
{2003A&A...404..593A,  
2005A&A...435..631A,   
2013A&A...555A..96S,   
2015A&A...576A...9A,   
2016A&A...588A..25M,   
2020A&A...635A.164S,   
2020A&A...635A.165C}.  
An interesting point for the present work is that  
{\tt  LPCODE} computes the white dwarf  
evolution  in a  self-consistent way, including the 
modifications in  the internal  
chemical distribution induced by dynamical convection, 
fingering convection, atomic diffusion, and nuclear reactions.
Atomic diffusion has been implemented following Burger's scheme 
\citep{1969fecg.book.....B} that provides the diffusion velocities 
in a multicomponent plasma under the influence of gravity, 
partial pressure, and induced electric fields. Partial ionization 
of metals is taken into account as it has important effects 
for the diffusion timescales.

We have introduced some changes in the code in order to 
simulate the accretion process. All computations were done by considering
accretion as a continuous process. Accreted material was assumed 
to be uniformly distributed on the star's surface (see below).  
We performed simulations for different accretion rates, ranging from 
$10^6$ g/s to $10^{12}$ g/s. Metal abundances of accreted matter were 
set to mimic the composition of the bulk Earth \citep{2001E&PSL.185...49A}.
Finally we used {\tt OPAL} radiative opacities for different 
metallicities \citep{1996ApJ...464..943I}, complemented with
the molecular opacities from \citep{1994ApJ...437..879A} at
low temperatures. Since bulk Earth composition was adopted for
the accreted matter, a new set of opacity tables was generated
from the OPAL website following this composition.

Considering that DA white dwarfs develop a convective envelope at effective
temperatures lower than about 15000 K
\citep{1998MNRAS.296..206A,2011MNRAS.413.2827C,2018ApJ...859L..19B}, 
we implemented the accretion
process in two different ways,  depending on the presence or not of
envelope convection\footnote{\cite{2019MNRAS.488.2503C} found 
in their 3D radiation hydrodynamics simulations that a superficial 
convection zone develops at $T_\mathrm{eff}\approx 18000$ K. 
Due to numerical stability considerations, particularly related 
to the treatment of diffusion, 1D evolutionary codes may undergo 
convergence difficulties  when the radially sampling extends 
too far out into the regions 
where 
convection first sets in, thus making it undetectable until
the instability penetrates deeper at lower effective temperatures. 
Thus it is not surprising that 3D hydrodynamics simulations find 
convection to start earlier, at 
higher effective temperatures, than 1D codes.}.  
For those models with a convective envelope, the
accreted material was instantaneously mixed in that region.  This is a
reasonable  approximation  since  the  convection  timescale  is  much
shorter  than the  evolutionary timescale 
\citep{2012A&A...539A..87V}.  Specifically, for  a given
integration  timestep, we  estimate the amount of  material that  is
accreted  according  to  the  accretion rate,  and  this  material  is
uniformly  distributed  in  the  whole  convective  zone. At  higher
effective  temperatures,  when  convection  is  absent,  a  different
criterion is instead  required to distribute the  accreted material in
the very outer layers of the  star. In this case, different approaches
have been used in the past, either based on some arbitrary selection
of  the depth  at  which  the accreted  matter  should be  distributed
homogeneously    \citep{2006A&A...453.1051K, 2009A&A...498..517K}
or  by considerations  that neglect the role of fingering convection 
instability \citep{2012MNRAS.424..333G}. It is worth noting that the 
deepening of the fingering convection instability 
eventually makes the choice of that depth less critical. We
performed additional  calculations to verify this. 

In this work, particular attention has been paid to the evolution
of  $^{16}$O, $^{24}$Mg, $^{28}$Si, $^{40}$Ca and  $^{56}$Fe. These
elements are important since they have been detected in the photospheres
of many white dwarfs 
\citep{2003ApJ...596..477Z,2007ApJ...671..872Z,2014ApJ...783...79X,
2019AJ....158..242X,2017ApJ...834....1M}. 
Their presence in the final surface composition of our simulations  allow 
us to link any given accretion rate with these surface abundances  for each  
model, characterized by its effective temperature and amount of
hydrogen.

The computation of energy transport was performed by using
the double diffusion theory of \cite{1993ApJ...407..284G}
as described by \cite{2011A&A...533A.139W}
Diffusion coefficients for fingering convection zones
were obtained by adopting the prescription of 
\cite{2013ApJ...768...34B}.

\section{Results}  
\label{results}
In this section we describe the main results of our simulations,
paying special attention not only to the final composition of the 
atmosphere but to the whole process that leads to the final state. 
Our models are characterized by three main parameters:
\begin{enumerate}
  \item the amount of hydrogen contained in the envelope ($M_\mathrm{H}$),
  \item the effective temperature ($T_\mathrm{eff}$),
  \item the accretion rate ($\dot M$).
\end{enumerate}
As mentioned before, four different models were considered, based on 
the amount of hydrogen that remains from the previous evolution. 
For each model we took initial configurations with ten different 
effective temperatures, ranging from 6000 K to 25000 K. Finally 
we subjected each model to four different accretion rates. The total 
number of simulations performed was 180. Table \ref{table:params} 
shows the details of the parameters adopted for each set of models.  
Some sets of parameters could not be combined to perform the
corresponding simulation because either the initial model 
was impossible to be generated as a hydrogen rich (DA) white dwarf,
or because a thin hydrogen envelope combined with a large accretion 
rate produced surface compositions running out of the
opacity tables.
\begin{table}
\caption{Adopted values for the set of parameters characterizing 
each model.}
\label{table:params}
\centering
\renewcommand{\arraystretch}{1.2}
\begin{tabular}{cc}
\hline\hline
 parameter & adopted values\\
\hline
$\log (M_\mathrm{H}/M_\odot)$ & -4, -6, -8, -10 \\
$\log \dot M$ & 6, 8, 10, 12\tablefootmark{a} \\
$T_\mathrm{eff}/10^3$K & 6\tablefootmark{b}, 8\tablefootmark{c}, 10, 10.5, 11, 11.5, 12, 16, 20, 25\\
\hline
\end{tabular}
\tablefoot{Some combinations of parameters have been discarded 
because of the following reasons:\\
\tablefoottext{a}{For maximum accretion rate, models with  
                  $\log (M_\mathrm{H}/M_\odot)= -10$ and 
                  $T_\mathrm{eff}\geq 10000$K run out of 
                  our opacity tables because of the excess 
                  of metal accumulation at the surface.}\\
\tablefoottext{b}{Transforms into a DB white dwarf for 
                  $\log (M_\mathrm{H}/M_\odot)= -8$ and $-10$.}\\
\tablefoottext{c}{Same as (b) but for $\log (M_\mathrm{H}/M_\odot)= -10$ 
                  models.}
}
\end{table}

\begin{figure}
\resizebox{\hsize}{!}{\includegraphics{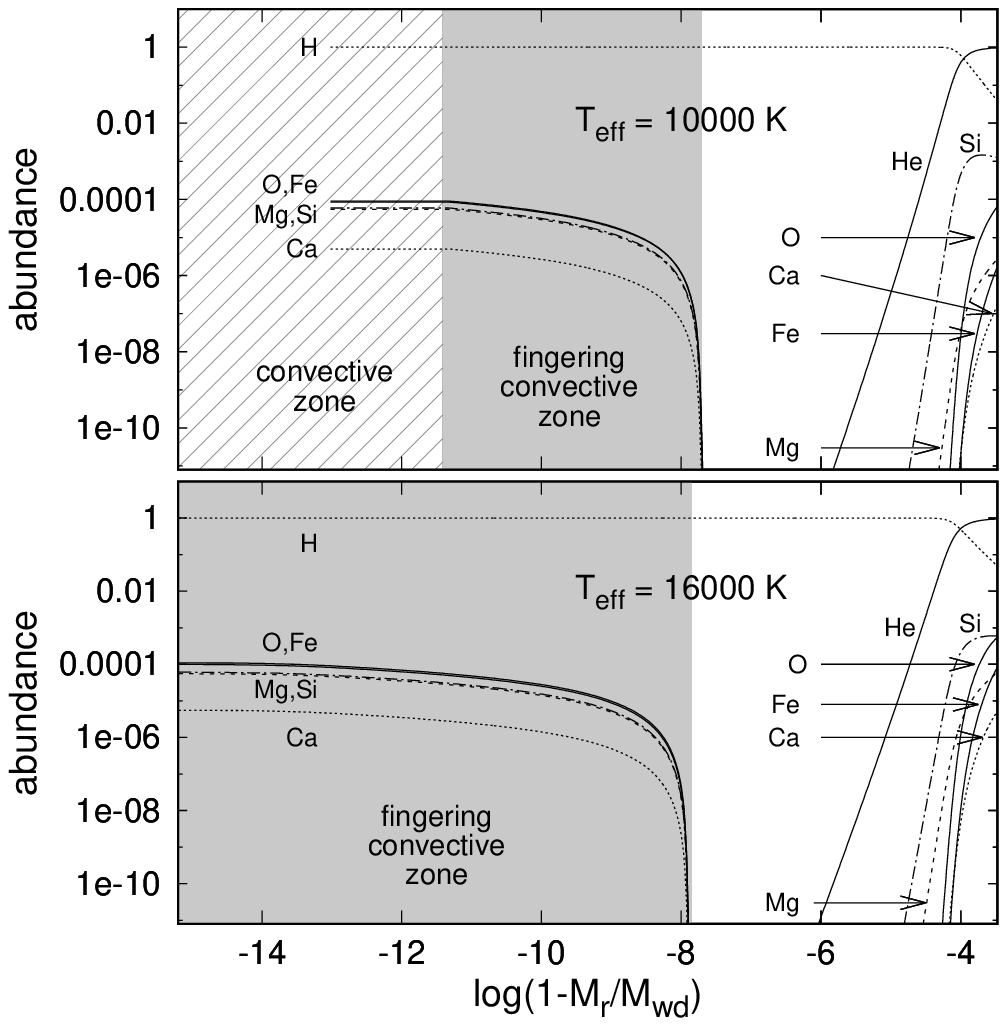}}
\caption{Final chemical profile for two models with 
$M_\mathrm{H}=10^{-4} M_\odot$ but different effective
temperatures. The accretion rate in both cases was of
$10^{10}$ g s$^{-1}$.}
\label{fig:MH4_CZyes_CZno}
\end{figure}

According to its effective temperature, a model may present a
convective envelope or not. This fact has some impact on the internal
structure  once  a  stationary  state  is reached.   For  instance,
Fig. \ref{fig:MH4_CZyes_CZno} shows the final chemical profile for two
models  with the  largest amount  of hydrogen  ($10^{-4}M_\odot$), one
having  a convective  envelope while  the other  not. Both  models were
obtained from an accretion rate of $10^{10}$ g/s.  Convection mixes up
the composition of  the superficial layers in a very short timescale,
thus leading  to a homogeneous abundance  of all elements in  that region
(shown in the  figure as a horizontal line in  the convective zone, CZ
hereafter).  Below the convective region  and because of the inversion
of the  molecular weight  ($\mu$), a  fingering convective  zone (FCZ)
sets in.  As we will discuss later,
the turbulent motions generated by this instability right
below  the   bottom  of  the   convective  zone  is   responsible  for
transporting the heavy elements coming from the upper layers down into
the deeper regions of the star. This figure clearly shows how the
turbulence in the FCZ diminishes as we go deeper. Indeed, the slope of
the chemical  profile of the heavy  elements in that region  goes from
almost horizontal  (more homogeneously distributed material)  near the
bottom  of  the  CZ,  to  very   steep  at  the  bottom  of  the  FCZ,
characterized  by  a  smooth  transition to  the  radiative  transport
regime. It is worth mentioning  that since fingering convection  is a
more efficient process than  element diffusion, larger accretion rates
are needed to maintain a given surface contamination when fingering
convection is taken into account, as shown by \cite{2013A&A...557L..12D} 
and \cite{2017A&A...601A..13W}.
We may show the incidence of taking fingering convection into
account by comparing the final (stationary)
surface abundance of one representative metal, namely iron,  
for simulations where we have turned on 
and off this process. Table \ref{table:MLTvsGNA} shows the results 
for these simulations performed using the same base model which has 
$M_\mathrm{H}=10^{-4} M_\odot $ and $T_\mathrm{eff}=10000$ K. From 
the table it becomes clear that fingering convection needs to be 
included when associating a surface contamination with the 
corresponding accretion rate. Neglecting this process 
results in superficial iron abundances which may differ by a 
factor up to 30 for these simulations.

\begin{table}
\caption{Final iron surface abundances (in mass) for simulations turning 
on and off fingering convection. Three accretion rates where
considered for initial models all having the same amount of hydrogen 
($M_\mathrm{H}=10^{-4} M_\odot $) and effective temperature 
($T_\mathrm{eff}=10000$ K). The last row shows the fractional 
difference between both models, with and without fingering convection.}

\label{table:MLTvsGNA}
\centering
\renewcommand{\arraystretch}{1.2}
\begin{tabular}{cccc}
\hline\hline
    & $10^6$ g/s & $10^8$ g/s & $10^{10}$ g/s \\
\hline
including FC & $0.80\times 10^{-7}$ & $0.35\times 10^{-5}$ & $0.90\times 10^{-4}$ \\
without FC   & $0.27\times 10^{-6}$ & $0.27\times 10^{-4}$ & $0.27\times 10^{-2}$ \\
without/with FC  & $3.4$  & $7.7$                &  $30$    \\
\hline
\end{tabular}
\end{table}

\begin{figure}
\resizebox{\hsize}{!}{\includegraphics{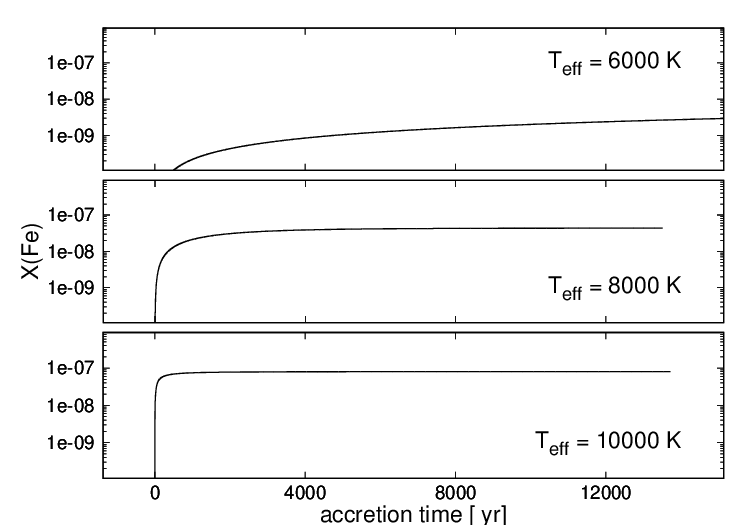}}
\caption{Temporal evolution of the abundance (in mass) of iron for a
  continuous accretion rate of $10^{6}$ g s$^{-1}$ for white dwarf models
  with a hydrogen content of $10^{-4} M_\odot$ at three selected
  effective temperatures.}
\label{fig:evol_abundances_FCZ}
\end{figure}

Fig. \ref{fig:evol_abundances_FCZ} displays  the temporal evolution of
the photosphere's abundance of iron for  an accretion rate of $10^6$ g
s$^{-1}$ at three  different effective temperatures. We  note that for
the hotter  models, those with $T_\mathrm{eff}=8000$ K  and 10000 K, the
abundance of  iron reaches a stationary  state well before the  end of
the  simulation,   which  was  set   after  14000  yr   of  continuous
accretion\footnote{The end of these simulations was 
arbitrarily set
to 14000 yr after confirming that a stationary state was reached.}.  
However, at  the lowest  effective temperature,  much more
time is required to reach the stationary state 
(about 200000 yr)\footnote{Some simulations took about one week to 
run and still did not reach a stationary state. In those cases we report 
a lower value for the particular element in the corresponding table.}. This
is an  expected behavior since as  CZ becomes more massive  as cooling
proceeds,  more  time is  needed  to  achieve the  final  (stationary)
state. We  mention that all of  our simulations have been  extended in
order to  reach a  final state  as close as  possible to  a stationary
situation.

\begin{figure}
\resizebox{\hsize}{!}{\includegraphics{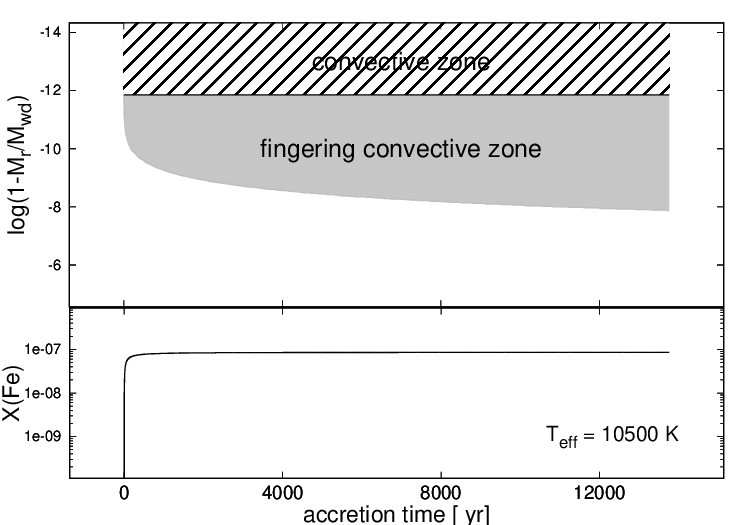}}
\caption{Upper panel: time evolution of the convective and 
fingering convective zone during the accretion period for
a model with $T_\mathrm{eff}=10500$ K, $M_\mathrm{H}= 10^{-4} M_\odot$
and accretion rate of $10^6$ g s$^{-1}$. Bottom panel: evolution 
of the iron abundance (in mass) at the surface.
}
\label{fig:evol_abundances_and_FCZ}
\end{figure}

\begin{figure}
\resizebox{\hsize}{!}{\includegraphics{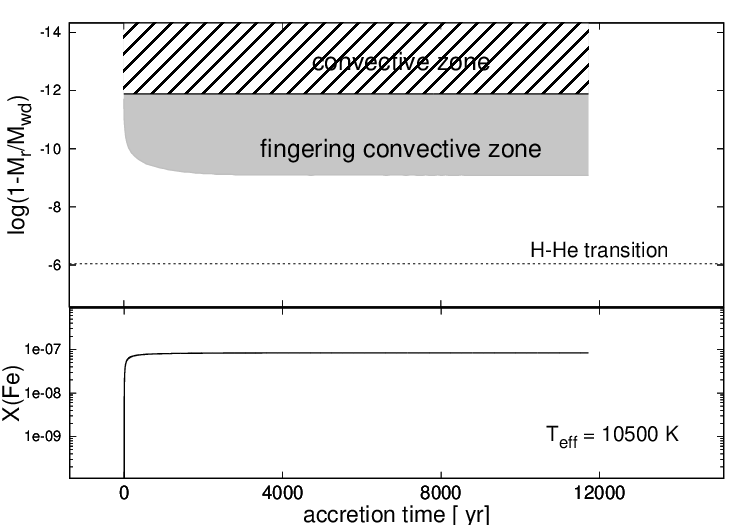}}
\caption{Same as Fig. \ref{fig:evol_abundances_and_FCZ} 
but for a model with $M_\mathrm{H}= 10^{-6} M_\odot$. A dotted
horizontal line shows the depth where hydrogen abundance by mass 
falls below 0.5.
}
\label{fig:evol_abundances_and_FCZ_2}
\end{figure}

Fig. \ref{fig:evol_abundances_and_FCZ} reveals  another feature of our
simulations, namely, the contrast between the time required for the
surface heavy elements to reach the stationary state and the evolution
of the FCZ.  Indeed, while the abundance of iron, as well as that of 
the other
heavy elements accreted (not shown), rapidly reach a stationary state,
the bottom of the FCZ continues moving to deeper layers during white
dwarf evolution. This is in contrast with the  situation shown  in
Fig. \ref{fig:evol_abundances_and_FCZ_2}  for the case of  a smaller H
envelope. Here, the inward advance of the bottom of the FCZ is halted
by the H-He transition, where the inverse  $\mu$-gradient produced by
the accretion is  counteracted by the strong chemical  gradient at the
H-He interface.  We note that in this case the  evolution of the iron
abundance is the same as that     shown      in
Fig.  \ref{fig:evol_abundances_and_FCZ}, i.e.  the  stationary state  is
reached in a short time.

\begin{figure}
\resizebox{\hsize}{!}{\includegraphics{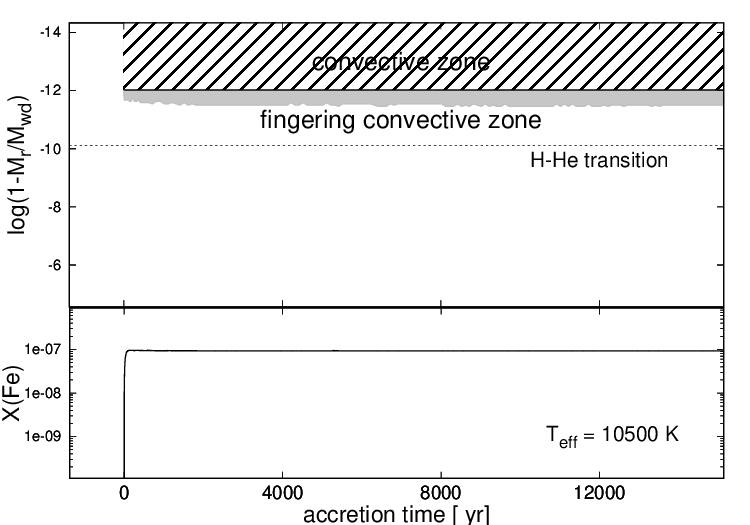}}
\caption{Same as Figure \ref{fig:evol_abundances_and_FCZ_2} 
but for a model with $M_\mathrm{H}= 10^{-10} M_\odot$.  
}
\label{fig:evol_abundances_and_FCZ_3}
\end{figure}

\begin{figure}
\resizebox{\hsize}{!}{\includegraphics{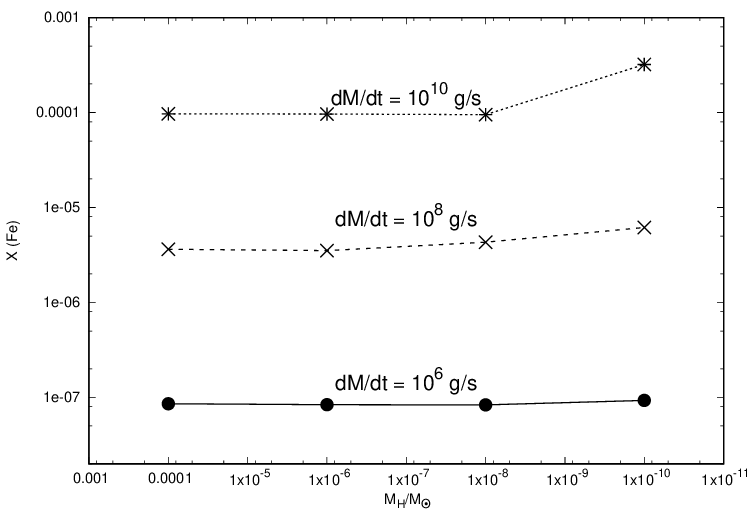}}
\caption{Final abundances (in mass) of iron for envelopes 
of different amounts of hydrogen and different accretion rates.
}
\label{fig:iron}
\end{figure}

The impact  of a thinner FCZ on the accumulation of heavy elements on
the  surface increases  when  the  H-He transition  is  closer to  the
photosphere  of the star.  Fig.   \ref{fig:evol_abundances_and_FCZ_3}
illustrates how  thin the FCZ  becomes when  the hydrogen mass  in the
envelope is  reduced to $10^{-10}  M_\odot$. In this case the 
bottom of the CZ penetrates more in the He-rich layers, lowering the 
contrast between the molecular weights inside the CZ and below. 
Because of the effect of the stabilizing $\mu$-gradient produced
by the increasing helium abundance as we go deeper, 
fingering convection barely shows up. 
Therefore, the expected presence of even a small amount of 
convective overshoot is likely to completely dominate such a small FCZ.
The extension of the FCZ
also depends on the accretion rate:  the higher the accretion rate the
wider the FCZ (not shown in the figure). In  the case shown  in Fig.
\ref{fig:evol_abundances_and_FCZ_3}, the  abundance of  iron increases
by  15\%   with  respect   to  the  cases   with  a   larger  hydrogen
envelope. 
This increase is not at all obvious in Fig. 
\ref{fig:evol_abundances_and_FCZ_3}, but 
is more evident for higher accretion rates. Fig. \ref{fig:iron} 
shows the dependence of the final 
iron abundances with the amount of hydrogen present in the 
envelope, for three different accretion rates. The maximum difference 
with respect to the case with larger hydrogen envelope 
happens for an accretion rate of $10^{10}$ g s$^{-1}$, being 
the increase of 233\%. In the intermediate case ($10^8$ g s$^{-1}$)  
the abundance increase is of 69\%. Other simulations show a much higher  
accumulation of heavy elements when the FCZ becomes thin.

Models with a larger content of hydrogen have the H-He transition deeper 
and that may cause this region to be unreachable for the FCZ. 
In fact, all our simulations using models with 
$M_\mathrm{H}=10^{-4} M_\odot$ show that the bottom of the 
FCZ does not reach the H-He transition layers. Thus, 
the FCZ finds no obstacle to advance deeper as the simulation
continues, although it slows down its pace as it penetrates 
into layers of increasing density. In contrast, the 
bottom of the convective zone (when the model has one) remains always 
at the same depth. Since the extension of the convective zone depends 
on the effective temperature, the level of accumulation of heavy elements
on the surface will also depend on this parameter. Cooler models, with 
larger convective zones, rapidly spread the accreted material into 
this larger region, producing less contamination of the surface than
in hotter white dwarfs. The evolution of the FCZ is faster for higher 
accretion rates, it also goes deeper carrying the heavy material further
inside the star. 

\begin{figure}
\resizebox{\hsize}{!}{\includegraphics{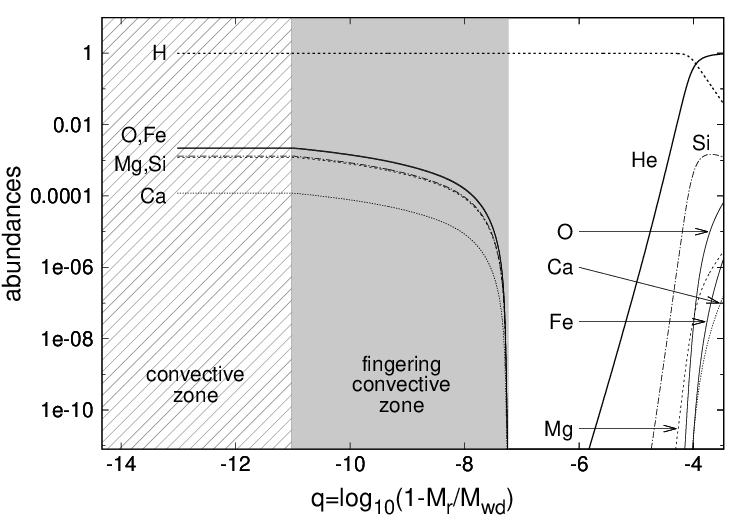}}
\caption{Chemical profile of the final configuration of a model with 
$10^{-4} M_\odot$ of hydrogen, $T_\mathrm{eff}=10000$ K and accretion
rate of $10^{12}$ g s$^{-1}$.  
}
\label{fig:chem_profile_10000K_acc12}
\end{figure}

Fig. \ref{fig:chem_profile_10000K_acc12} shows the 
chemical profile at the end of the simulation for a model with 
$M_\mathrm{H}=10^{-4} M_\odot$, $T_\mathrm{eff}=10000$ K and maximum 
accretion rate ($10^{12}$ g s$^{-1}$). The FCZ 
extends through a large
region of the star (in a logarithmic scale in mass) but is still far
from reaching the He-rich layers. Accreted heavy elements are homogeneously
distributed throughout the convective zone but are less and less abundant as we 
advance deeper through the FCZ. The turbulence associated to the fingering
convective instability is maximum near the bottom of the convective zone
and diminishes downwards, tending gradually to zero.
In our 1D computations, there is
a somewhat sharp step in turbulence between dynamical and fingering 
convective zones, which would be smoother if 3D simulations of convection 
were taken into account 
\citep{1996A&A...313..497F,2018MNRAS.474.4660K,2019MNRAS.488.2503C}. 
We did not add any 1D parametrization of overshoot or penetrative convection. 
A rapidly decreasing extra-mixing below the CZ would smoothen the local 
$\mu$-gradient at the beginning of the simulations. Fingering convection 
would rapidly takeover, leading to similar final results. 
Detailed computations with various parameterizations of the bottom 
of the convective zones may be undertaken in the future.


In the case of thinner envelopes, the chemical evolution of 
accreting white dwarfs is quite different from the 
$M_\mathrm{H}=10^{-4} M_\odot$ case described 
before\footnote{\cite{2020MNRAS.492.3540C} 
found observational evidence that approximately 20\% of 
white dwarfs are expected to have a hydrogen
content of $-14 < \log (M_\mathrm{H}/M_\mathrm{wd})<-10$
whilst approximately 65\% have  
$\log (M_\mathrm{H}/M_\mathrm{wd}) > -10$.}.
The main reason for such a difference is that 
now the turbulence from the upper layers is able to reach 
the H-He transition zone, something that does not happen for
thicker envelopes. 

\begin{figure}
\resizebox{\hsize}{!}{\includegraphics{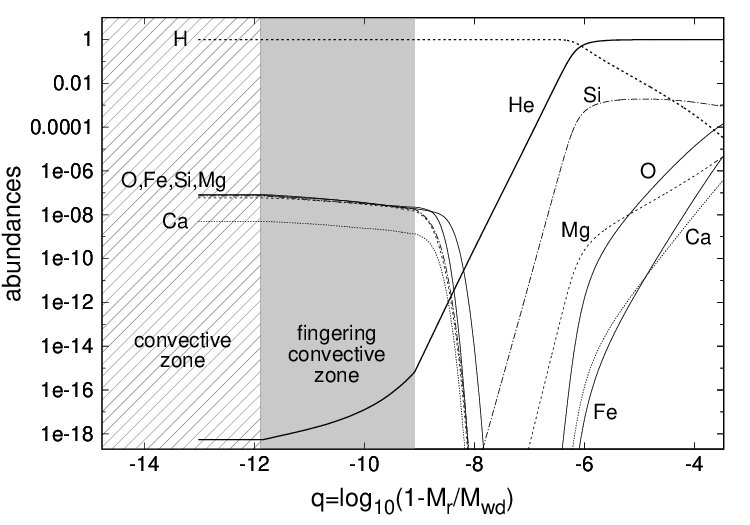}}
\caption{Chemical profile of the final configuration of a model with 
$10^{-6} M_\odot$ of hydrogen, $T_\mathrm{eff}=10500$ K and accretion
rate of $10^{6}$ g s$^{-1}$.  
}
\label{fig:chem_profile_10500K_acc6}
\end{figure}

We may start by describing our results for models with 
$M_\mathrm{H}=10^{-6} M_\odot$. For such a thin envelope, 
the FCZ which develops below the CZ expands until penetrating
the transition zone where He becomes more abundant\footnote{
We found only two cases where no FCZ was formed. Both 
simulations correspond to the coolest models 
($T_\mathrm{eff}=6000$ K) and to the lower accretion rates ($10^6$ 
and $10^8$ g s$^{-1}$). Two related facts are responsible for this
result. First, the bottom of the 
convective zone deeply penetrates the H-He transition region 
reaching a stabilizing $\mu$-gradient due to the increasing
abundance of He as we go deeper. Second, the low accretion rate 
combined with a large CZ (which strongly dilutes the abundance 
of heavy elements), lowers the difference between the
molecular weights at the bottom of the CZ and the region 
immediately below it.  
The combination of these factors precludes the formation of the FCZ.}. 
This encounter prevents the FCZ to go deeper, as it is stabilized by the normal 
$\mu$-gradient due to the increasing amount of He. Thus, the heavy elements 
accumulated in the FCZ continue to progress further down by diffusion 
in a radiative medium, something that never happened in our 
previously described simulations of models with thicker envelopes, 
since the presence of heavy elements in a 
H-rich medium always triggered the fingering convection instability. 
Figure \ref{fig:chem_profile_10500K_acc6} 
shows such a situation for a model of $T_\mathrm{eff}=10500$ K and 
an accretion rate of $10^6$ g s$^{-1}$. We expanded the abundance 
range to include very low values in order to
show the contact between the bottom of the FCZ and the He tail, which stops
the instability. We also note the dredge-up of He by the FCZ, which leads
to the contamination of the surface by a very small amount in this case. 
As can be seen from Figure \ref{fig:chem_profile_10500K_acc6}, 
the FCZ stops where the He/H abundance ratio reaches approximately $10^{-15}$. 
The consecutive dredge-up of He would lead to an undetectable He abundance 
of $10^{-18.87}$ in the photosphere (see Table \ref{table:3}).
 
\begin{figure}
\resizebox{\hsize}{!}{\includegraphics{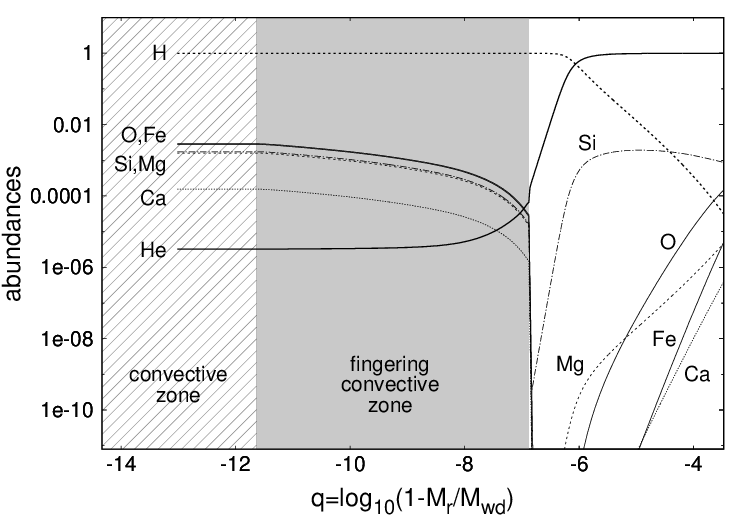}}
\caption{Same as Fig. \ref{fig:chem_profile_10500K_acc6} but for a model with 
$10^{-6} M_\odot$ of hydrogen, $T_\mathrm{eff}=10500$ K but accretion
rate of $10^{12}$ g s$^{-1}$.  
}
\label{fig:chem_profile_10500K_acc12}
\end{figure}

Larger He contamination is expected in the case of larger
accretion rates, see  Fig. \ref{fig:chem_profile_10500K_acc12}. In this
case, the FCZ is able to further penetrate the H-He transition region,
with the consequent larger He enrichment of the outer layers. 
One can see from Fig. \ref{fig:chem_profile_10500K_acc12} 
that for an accretion rate of $10^{12}$ g s$^{-1}$, the FCZ stops where 
the He/H abundance ratio reaches approximately $10^{-4}$. 
In this case, our simulation was interrupted before the steady state 
for the photospheric He abundance could be reached.  
The achieved lower limit for the abundance of He in the 
photosphere is $10^{-6.09}$ (see Table \ref{table:3}). 
The photospheric He abundance depends on both the hydrogen mass 
fraction and the accretion rate. In the cases where the hydrogen mass 
fraction could be independently derived and the accretion rate estimated 
from the observed heavy elements abundances, it should be possible to 
distinguish whether the photospheric helium has been accreted or dredged-up.

All the simulations
with the highest accretion rates show this kind of behavior.
Unfortunately the abundance of He on the surface takes much longer to reach
a steady state than the accreted heavy elements, and we had to stop the 
simulations before that state was reached because of the excessive time required by the 
computation. We estimate that it takes of the order of 15000 yr of 
evolution to finally reach that steady state.
Thus, our tabulated abundances of helium are only lower boundaries in most cases.

About 25\% of the parameter space covered in this work was 
studied already by \cite{2019ApJ...872...96B} for
a $\log(M_\mathrm{H}/M_\mathrm{wd})=-6$ model. 
Table \ref{table:compare_BB19} compares both results, adding those 
obtained by \cite{2009A&A...498..517K}. In all cases we show the 
diffusion calculations using the coefficients of \cite{1986ApJS...61..177P}. 
Some differences in the convective zone are apparent, 
arising from the differences between the Mixing Length Theory used 
by those works and the \cite{1993ApJ...407..284G} convection theory implemented
here.

Models with $M_\mathrm{H}=10^{-8} M_\odot$ share many of the main 
features of the $M_\mathrm{H}=10^{-6} M_\odot$ case. 
Dredge-up of helium is much more efficient now  
and consequently the contamination of 
the surface by helium is noticeably higher. The thinnest envelope 
models ($M_\mathrm{H}=10^{-10} M_\odot$) continue with the
same tendency: more helium contamination of the surface 
and a FCZ advance stopped earlier by the more superficial H-He transition
zone.

\begin{figure*}
\centering\includegraphics[width=16cm]
{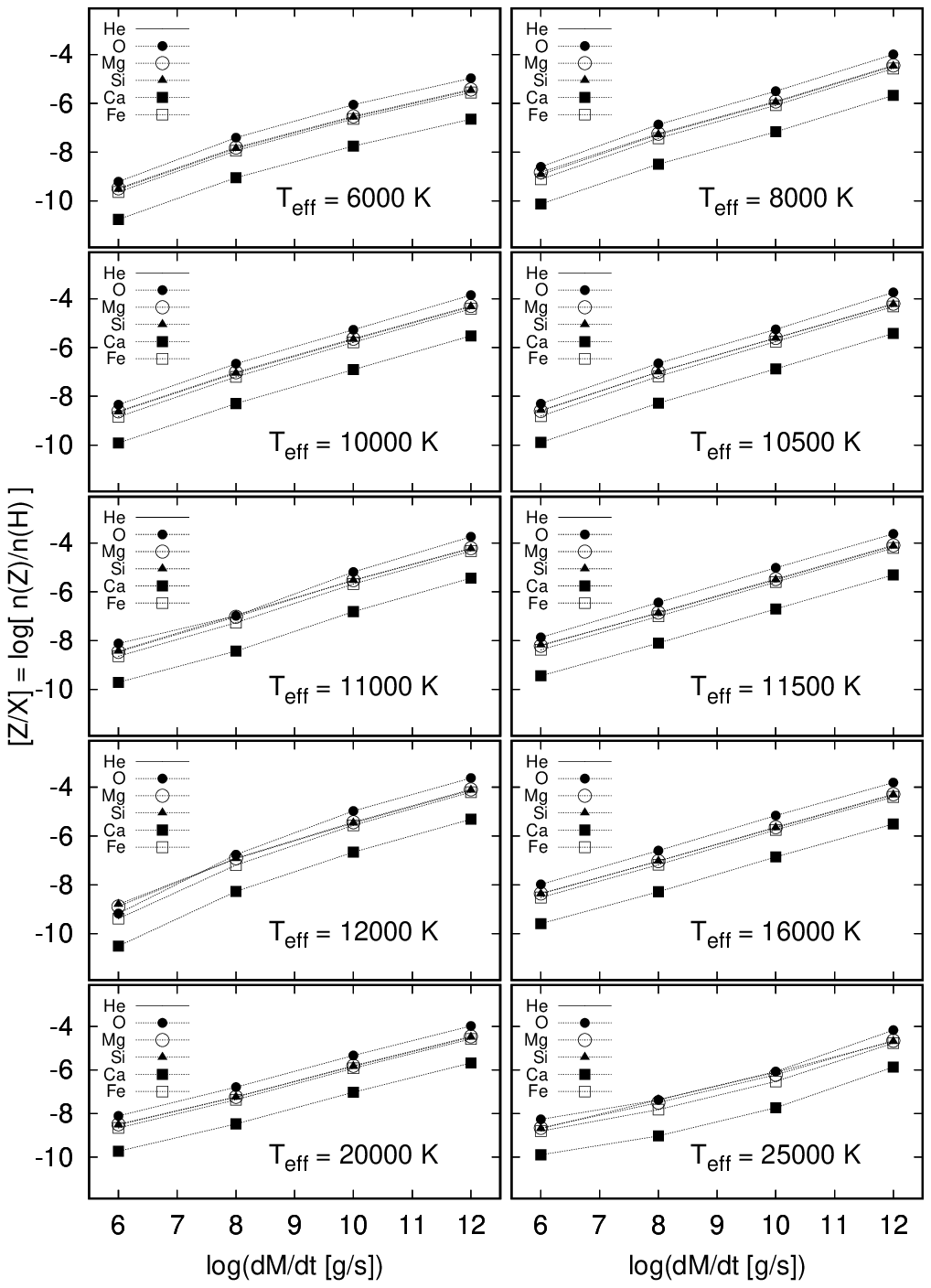}
\caption{Surface contamination against the accretion rate for 
 models with $M_\mathrm{H}=10^{-4} M_\odot$. The contamination
is given in terms of the abundance expressed in [Z/H]$=\log n(\mathrm{Z})/n(\mathrm{H})$.}
\label{fig:surf_chem_profile_4}
\end{figure*}

\begin{figure*}
\centering\includegraphics[width=16cm]{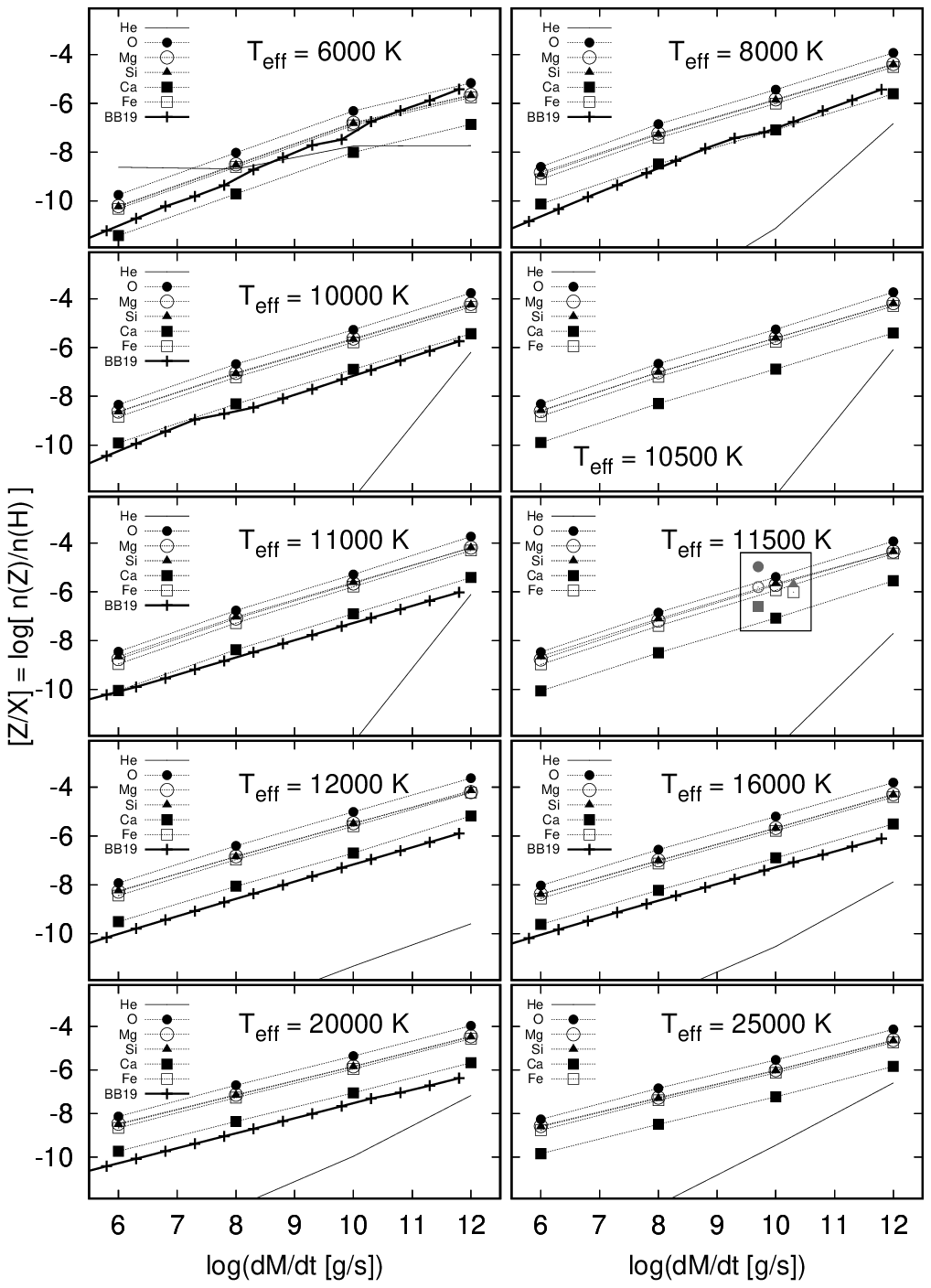}
\caption{Same as Figure \ref{fig:surf_chem_profile_4} 
for models with $M_\mathrm{H}=10^{-6} M_\odot$. When possible, $^{40}$Ca 
abundances obtained by  \cite{2019ApJ...872...96B} have been added to 
the corresponding panel (labeled as BB19). For $T_\mathrm{eff}=11500$ K, 
gray symbols (inside a box) represent the abundances obtained by \cite{2018ApJ...859L..19B} 
for $^{16}$O, $^{24}$Mg, $^{28}$Si, $^{40}$Ca and $^{56}$Fe using the 
observed photospheric abundance of pollutants in G29--38. All these isolated points correspond to 
an accretion rate of $10^{10}$ g s$^{-1}$ but have been
shifted a bit from that particular value for the sake of clarity.}
\label{fig:surf_chem_profile_6}
\end{figure*}

\begin{figure*}
\centering\includegraphics[width=16cm]{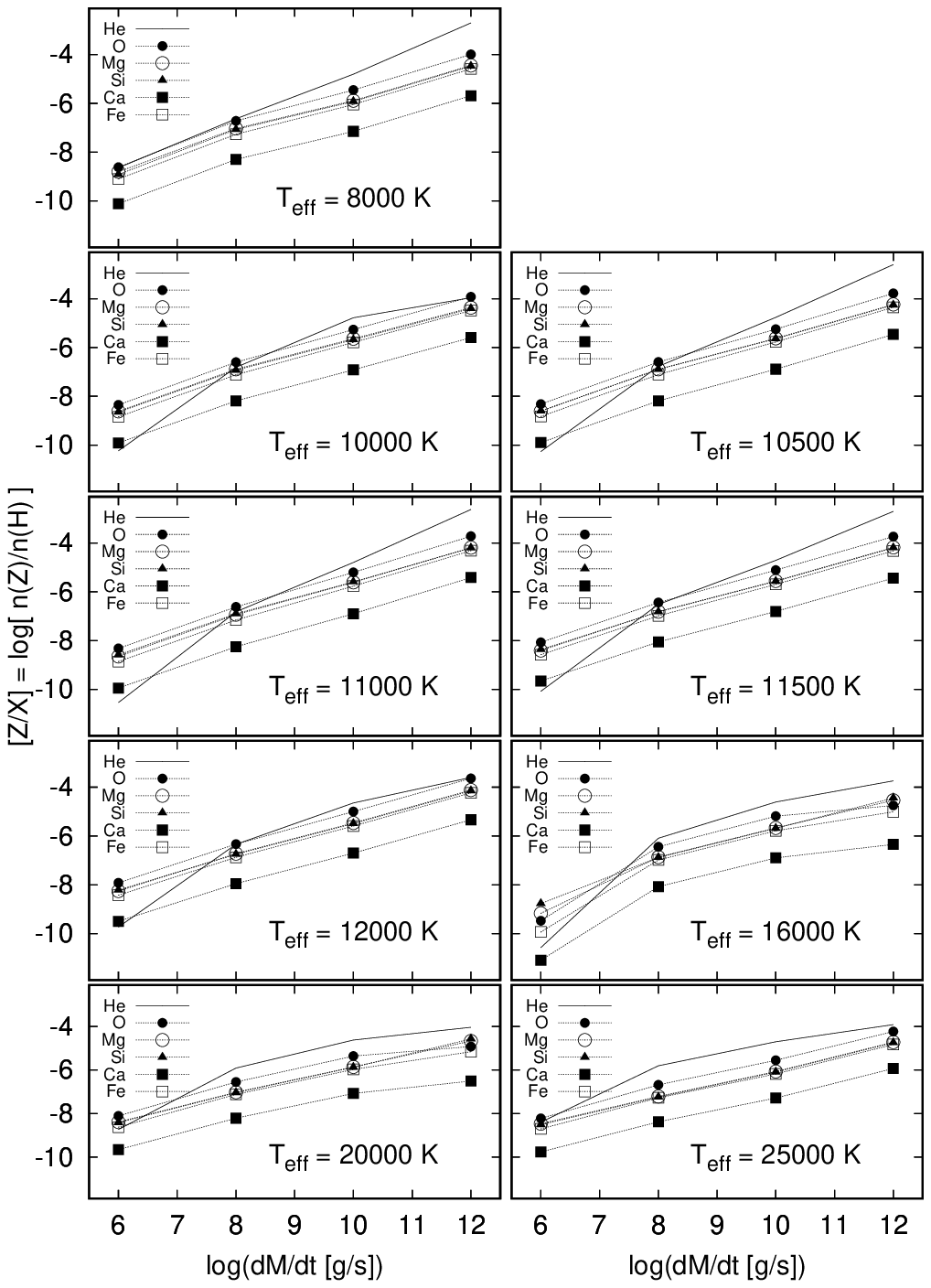}
\caption{Same as Figure \ref{fig:surf_chem_profile_4} 
for models with $M_\mathrm{H}=10^{-8} M_\odot$.}
\label{fig:surf_chem_profile_8}
\end{figure*}

\begin{figure*}
\centering\includegraphics[width=16cm]{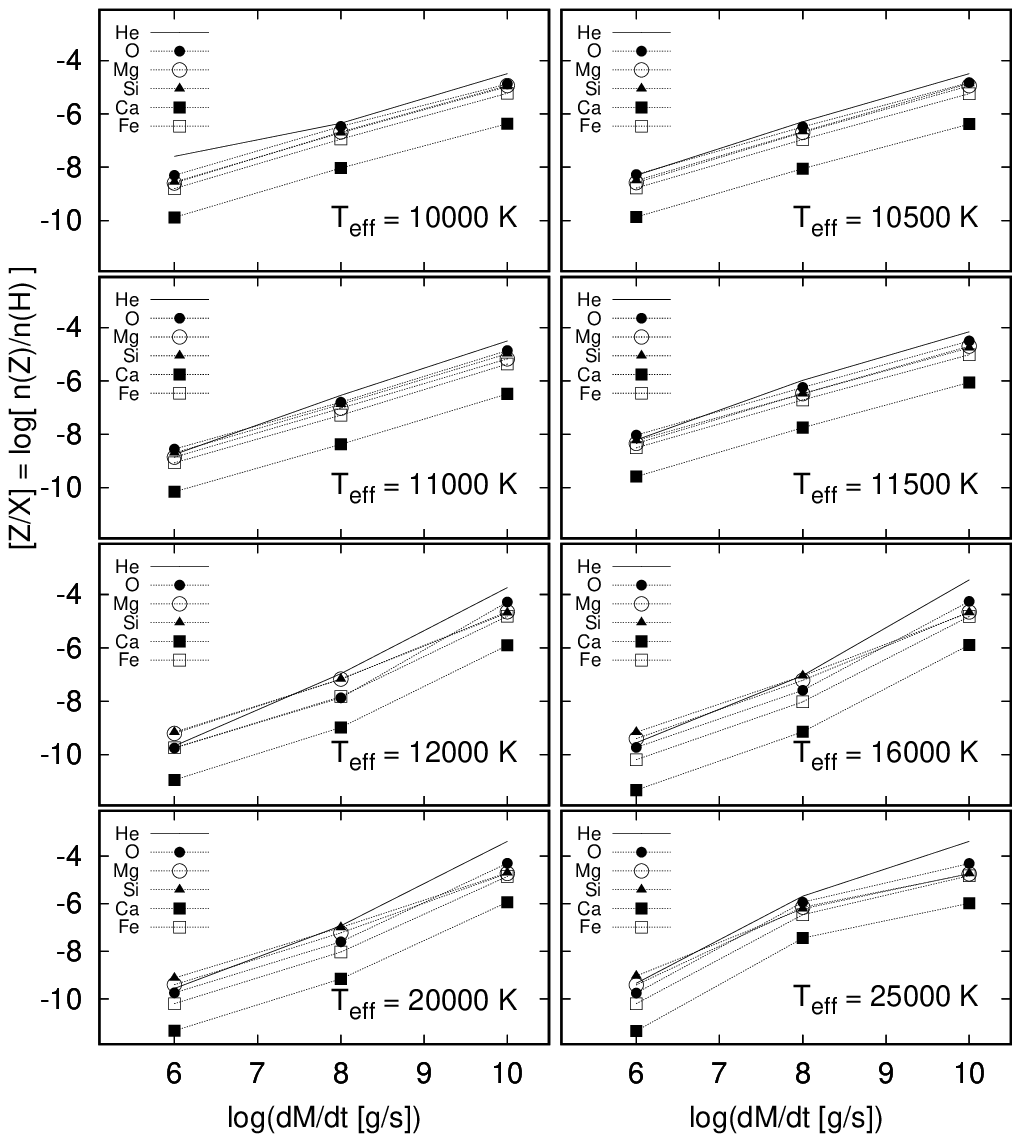}
\caption{Same as Figure \ref{fig:surf_chem_profile_4} 
for models with $M_\mathrm{H}=10^{-10} M_\odot$.}
\label{fig:surf_chem_profile_10}
\end{figure*}

Figures \ref{fig:surf_chem_profile_4} to \ref{fig:surf_chem_profile_10}
show how the surface contamination changes with the accretion rate for
models of four different hydrogen envelopes. The details of the abundances
at the end of each simulation are summarized 
in Tables \ref{table:2} to \ref{table:4}.

Figure \ref{fig:surf_chem_profile_6} includes  the
surface mass fractions for $^{40}$Ca obtained by 
\cite{2019ApJ...872...96B}. Although not strictly the
same temperature, we include also their results
for $T_\mathrm{eff}=15000$ K and 20500 K in our panels for 16000 K and
20000 K, respectively. They are some systematic differences for higher 
temperatures, where we obtain somewhat higher $^{40}$Ca abundances than 
\cite{2019ApJ...872...96B}. A better  agreement is found for  
$T_\mathrm{eff} \leq 10000$ K, although for $T_\mathrm{eff}=6000$ K 
we obtain smaller abundances for higher accretion rates. This difference 
is due to difficulties to reach the final steady state in our simulations 
because of the small timestep needed by our diffusion calculation 
(in these cases) to fulfill the required precision. Therefore, our results 
for $T_\mathrm{eff}=6000$ K and high accretion rates should be taken as 
lower boundaries. Figure \ref{fig:surf_chem_profile_6}
also includes for $T_\mathrm{eff}=11500$ K the steady state abundances 
for $^{16}$O, $^{24}$Mg, $^{28}$Si, $^{40}$Ca and $^{56}$Fe provided by 
\cite{2018ApJ...859L..19B} using the observed photospheric abundance of 
pollutants in G29--38. There is a good agreement for the abundances of
$^{24}$Mg, $^{28}$Si and $^{56}$Fe, whereas $^{16}$O and $^{40}$Ca show 
higher values (by about  $\Delta [Z/X]=0.44$) in 
\cite{2018ApJ...859L..19B} work.

\begin{table*}   
\caption{Comparison of \cite{2009A&A...498..517K},
\cite{2019ApJ...872...96B} and our results for the
mass of the surface convection zone and diffusion 
timescales for $^{40}$Ca on a 0.6 $M_\odot$ white
dwarf.}
\label{table:compare_BB19}
\centering
\begin{tabular}{cccccccccc}
\hline
    $T_\mathrm{eff}$  &           & $\log(M_\mathrm{cvz}/M_\mathrm{wd})$ &             & \multicolumn{2}{c}{$\log g$}  &         & $\log(\tau_\mathrm{diff}/\mathrm{yr})$  \\                       
 \cline{2-4}
 \cline{7-9}
                      & Koester   &     BB19                 &   here      & BB19     &   here      & Koester &  BB19   & here \\                       
\hline
      6000 K          & $-7.722$  &  $-7.8094$               &  $-7.5803$ & $8.0342$ & $8.0594 $    & $4.2924$  & $4.2449$  & $4.4526$ \\
      8000 K          & $-8.432$  &  $-8.9849$               &  $-9.7887$ & $8.0272$ & $8.0515 $    & $3.3303$  & $3.4113$  & $3.0643$ \\
     10000 K          & $-10.738$ &  $-10.251$               &  $-11.456$ & $8.0202$ & $8.0447 $    & $1.9997$  & $2.476$   & $1.6092$ \\
     11000 K          & $-12.715$ &  $-11.872$               &  $-12.714$ & $8.0164$ & $8.0413 $    & $0.4845$  & $1.1984$  & $0.2165$ \\
     12000 K          & $-15.618$ &  $-14.698$               &  $-14.994$ & $8.0127$ & $8.0378 $    & $-1.6941$ & $-1.0767$ & $-1.5267$ \\
\hline
\end{tabular}
\end{table*}

\begin{table}   
\caption{Models with $M_\mathrm{H}=10^{-4} M_\odot$. Abundances are
expressed in [Z/H], where [Z/H]$=\log n(\mathrm{Z})/n(\mathrm{H})$ is 
the logarithmic number ratio of the abundance (in number) of element Z relative
to the abundance of hydrogen (H). Lower values are marked by a $>$ sign, when
the stationary state was not achieved and the surface abundance of that
element was still increasing.}
\label{table:2}
\centering
\begin{tabular}{cccccc}
\hline
            &     & 10$^6$ g s$^{-1}$ & 10$^8$ g s$^{-1}$ & 10$^{10}$ g s$^{-1}$  & 10$^{12}$ g s$^{-1}$\\
\hline
            & He & $ -20.34$ & $-19.67$ & $-19.15$ & $-19.25$ \\
            &  O & $  -9.16$ & $ -7.29$ & $ -5.88$ & $ -4.73$ \\
            & Mg & $  -9.31$ & $ -7.62$ & $ -6.34$ & $ -5.16$ \\
     6000 K & Si & $  -9.35$ & $ -7.66$ & $ -6.38$ & $ -5.21$ \\
            & Ca & $ -10.63$ & $ -8.88$ & $ -7.56$ & $ -6.38$ \\
            & Fe & $  -9.50$ & $ -7.76$ & $ -6.45$ & $ -5.30$ \\
\hline
            & He & $ -29.63$ & $-26.03$ & $-25.35$ & $-21.38$ \\
            &  O & $  -8.60$ & $ -6.86$ & $ -5.50$ & $ -3.99$ \\
            & Mg & $  -8.83$ & $ -7.24$ & $ -5.92$ & $ -4.44$ \\
     8000 K & Si & $  -8.91$ & $ -7.29$ & $ -5.95$ & $ -4.47$ \\
            & Ca & $ -10.12$ & $ -8.49$ & $ -7.16$ & $ -5.67$ \\
            & Fe & $  -9.11$ & $ -7.43$ & $ -6.07$ & $ -4.57$ \\
\hline
            & He & $ -29.84$ & $-25.00$ & $-24.15$ & $-21.93$ \\
            &  O & $  -8.34$ & $ -6.67$ & $ -5.28$ & $ -3.85$ \\
            & Mg & $  -8.62$ & $ -7.02$ & $ -5.64$ & $ -4.29$ \\
    10000 K & Si & $  -8.64$ & $ -7.05$ & $ -5.67$ & $ -4.33$ \\
            & Ca & $  -9.91$ & $ -8.30$ & $ -6.90$ & $ -5.52$ \\
            & Fe & $  -8.84$ & $ -7.21$ & $ -5.79$ & $ -4.41$ \\
\hline
            & He & $ -29.61$ & $-24.73$ & $-23.87$ & $-19.56$ \\
            &  O & $  -8.30$ & $ -6.65$ & $ -5.25$ & $ -3.74$ \\
            & Mg & $  -8.60$ & $ -7.00$ & $ -5.61$ & $ -4.20$ \\
    10500 K & Si & $  -8.58$ & $ -7.00$ & $ -5.62$ & $ -4.23$ \\
            & Ca & $  -9.89$ & $ -8.28$ & $ -6.88$ & $ -5.42$ \\
            & Fe & $  -8.81$ & $ -7.19$ & $ -5.76$ & $ -4.32$ \\
\hline
            & He & $ -28.54$ & $-24.79$ & $-21.93$ & $-23.69$ \\
            &  O & $  -8.12$ & $ -6.98$ & $ -5.18$ & $ -3.74$ \\
            & Mg & $  -8.46$ & $ -7.03$ & $ -5.53$ & $ -4.21$ \\
    11000 K & Si & $  -8.43$ & $ -6.96$ & $ -5.53$ & $ -4.24$ \\
            & Ca & $  -9.72$ & $ -8.43$ & $ -6.81$ & $ -5.44$ \\
            & Fe & $  -8.64$ & $ -7.27$ & $ -5.68$ & $ -4.33$ \\
\hline
            & He & $ -27.43$ & $-29.25$ & $-27.12$ & $-23.95$ \\
            &  O & $  -7.87$ & $ -6.44$ & $ -5.01$ & $ -3.62$ \\
            & Mg & $  -8.21$ & $ -6.85$ & $ -5.48$ & $ -4.09$ \\
    11500 K & Si & $  -8.18$ & $ -6.88$ & $ -5.51$ & $ -4.12$ \\
            & Ca & $  -9.45$ & $ -8.10$ & $ -6.70$ & $ -5.31$ \\
            & Fe & $  -8.37$ & $ -7.01$ & $ -5.60$ & $ -4.21$ \\
\hline
            & He & $ -28.69$ & $-29.32$ & $-25.13$ & $-24.59$ \\
            &  O & $  -9.18$ & $ -6.76$ & $ -4.97$ & $ -3.63$ \\
            & Mg & $  -8.89$ & $ -6.91$ & $ -5.44$ & $ -4.09$ \\
    12000 K & Si & $  -8.80$ & $ -6.92$ & $ -5.46$ & $ -4.12$ \\
            & Ca & $ -10.51$ & $ -8.27$ & $ -6.66$ & $ -5.31$ \\
            & Fe & $  -9.39$ & $ -7.19$ & $ -5.56$ & $ -4.20$ \\
\hline
            & He & $ -25.29$ & $-27.79$ & $-23.90$ & $-23.41$ \\
            &  O & $  -7.98$ & $ -6.60$ & $ -5.16$ & $ -3.81$ \\
            & Mg & $  -8.35$ & $ -7.03$ & $ -5.63$ & $ -4.29$ \\
    16000 K & Si & $  -8.37$ & $ -7.03$ & $ -5.66$ & $ -4.32$ \\
            & Ca & $  -9.59$ & $ -8.28$ & $ -6.86$ & $ -5.50$ \\
            & Fe & $  -8.52$ & $ -7.17$ & $ -5.75$ & $ -4.40$ \\
\hline
            & He & $ -18.27$ & $-29.02$ & $-22.92$ & $-22.45$ \\
            &  O & $  -8.11$ & $ -6.78$ & $ -5.33$ & $ -3.98$ \\
            & Mg & $  -8.49$ & $ -7.23$ & $ -5.81$ & $ -4.46$ \\
    20000 K & Si & $  -8.52$ & $ -7.24$ & $ -5.84$ & $ -4.49$ \\
            & Ca & $  -9.73$ & $ -8.47$ & $ -7.03$ & $ -5.68$ \\
            & Fe & $  -8.66$ & $ -7.36$ & $ -5.92$ & $ -4.57$ \\
\hline
            & He & $ -22.46$ & $-25.57$ & $-21.90$ & $-21.19$ \\
            &  O & $  -8.26$ & $ -7.37$ & $ -6.07$ & $ -4.16$ \\
            & Mg & $  -8.66$ & $ -7.49$ & $ -6.21$ & $ -4.65$ \\
    25000 K & Si & $  -8.69$ & $ -7.36$ & $ -6.10$ & $ -4.68$ \\
            & Ca & $  -9.90$ & $ -9.03$ & $ -7.73$ & $ -5.86$ \\
            & Fe & $  -8.81$ & $ -7.82$ & $ -6.52$ & $ -4.76$ \\
\hline
\end{tabular}
\end{table}

\begin{table}   
\caption{Same as Table \ref{table:2} for models with $M_\mathrm{H}=10^{-6} M_\odot$}
\label{table:3}
\centering
\begin{tabular}{cccccc}
\hline
            &     & 10$^6$ g s$^{-1}$ & 10$^8$ g s$^{-1}$ & 10$^{10}$ g s$^{-1}$  & 10$^{12}$ g s$^{-1}$\\
\hline
            & He & $  -8.50$ & $ -8.33$ & $ -7.75$ & $ >-7.74$ \\
            &  O & $  -9.34$ & $ -7.51$ & $ -6.31$ & $ -5.16$ \\
            & Mg & $  -9.71$ & $ -7.94$ & $ -6.79$ & $ -5.65$ \\
     6000 K & Si & $  -9.74$ & $ -7.97$ & $ -6.83$ & $ -5.68$ \\
            & Ca & $ -10.95$ & $ -9.15$ & $ -8.01$ & $ -6.86$ \\
            & Fe & $  -9.83$ & $ -8.04$ & $ -6.90$ & $ -5.75$ \\
\hline
            & He & $ -17.79$ & $-14.07$ & $-11.12$ & $ >-6.83$ \\
            &  O & $  -8.61$ & $ -6.85$ & $ -5.44$ & $ -3.92$ \\
            & Mg & $  -8.83$ & $ -7.23$ & $ -5.83$ & $ -4.38$ \\
     8000 K & Si & $  -8.91$ & $ -7.27$ & $ -5.87$ & $ -4.41$ \\
            & Ca & $ -10.13$ & $ -8.48$ & $ -7.09$ & $ -5.61$ \\
            & Fe & $  -9.11$ & $ -7.42$ & $ -6.00$ & $ -4.51$ \\
\hline
            & He & $ -18.86$ & $-14.71$ & $-12.18$ & $ >-6.20$ \\
            &  O & $  -8.34$ & $ -6.68$ & $ -5.27$ & $ -3.76$ \\
            & Mg & $  -8.62$ & $ -7.04$ & $ -5.63$ & $ -4.21$ \\
    10000 K & Si & $  -8.64$ & $ -7.07$ & $ -5.67$ & $ -4.25$ \\
            & Ca & $  -9.91$ & $ -8.31$ & $ -6.90$ & $ -5.44$ \\
            & Fe & $  -8.84$ & $ -7.22$ & $ -5.79$ & $ -4.33$ \\
\hline
            & He & $ -18.87$ & $-14.64$ & $-12.05$ & $ >-6.09$ \\
            &  O & $  -8.31$ & $ -6.66$ & $ -5.26$ & $ -3.73$ \\
            & Mg & $  -8.61$ & $ -7.02$ & $ -5.61$ & $ -4.18$ \\
    10500 K & Si & $  -8.59$ & $ -7.02$ & $ -5.62$ & $ -4.20$ \\
            & Ca & $  -9.90$ & $ -8.30$ & $ -6.88$ & $ -5.41$ \\
            & Fe & $  -8.82$ & $ -7.20$ & $ -5.76$ & $ -4.30$ \\
\hline
            & He & $ -19.49$ & $-14.30$ & $-12.17$ & $ >-6.11$ \\
            &  O & $  -8.45$ & $ -6.77$ & $ -5.28$ & $ -3.73$ \\
            & Mg & $  -8.76$ & $ -7.10$ & $ -5.65$ & $ -4.19$ \\
    11000 K & Si & $  -8.66$ & $ -7.03$ & $ -5.62$ & $ -4.20$ \\
            & Ca & $ -10.05$ & $ -8.38$ & $ -6.90$ & $ -5.41$ \\
            & Fe & $  -8.96$ & $ -7.29$ & $ -5.79$ & $ -4.30$ \\
\hline
            & He & $ -19.51$ & $-14.52$ & $-12.42$ & $ >-7.70$ \\
            &  O & $  -8.47$ & $ -6.85$ & $ -5.39$ & $ -3.93$ \\
            & Mg & $  -8.77$ & $ -7.18$ & $ -5.72$ & $ -4.35$ \\
    11500 K & Si & $  -8.64$ & $ -7.11$ & $ -5.66$ & $ -4.35$ \\
            & Ca & $ -10.06$ & $ -8.50$ & $ -7.08$ & $ -5.54$ \\
            & Fe & $  -8.97$ & $ -7.40$ & $ -5.94$ & $ -4.43$ \\
\hline
            & He & $ -19.30$ & $-13.16$ & $-11.33$ & $ >-8.20$ \\
            &  O & $  -7.93$ & $ -6.40$ & $ -5.01$ & $ -3.61$ \\
            & Mg & $  -8.28$ & $ -6.83$ & $ -5.48$ & $ -4.09$ \\
    12000 K & Si & $  -8.24$ & $ -6.85$ & $ -5.50$ & $ -4.11$ \\
            & Ca & $  -9.51$ & $ -8.05$ & $ -6.70$ & $ -5.31$ \\
            & Fe & $  -8.44$ & $ -6.97$ & $ -5.60$ & $ -4.20$ \\
\hline
            & He & $ -18.27$ & $-12.62$ & $-10.52$ & $ >-7.88$ \\
            &  O & $  -8.03$ & $ -6.56$ & $ -5.20$ & $ -3.81$ \\
            & Mg & $  -8.38$ & $ -7.00$ & $ -5.67$ & $ -4.29$ \\
    16000 K & Si & $  -8.39$ & $ -7.02$ & $ -5.69$ & $ -4.32$ \\
            & Ca & $  -9.62$ & $ -8.22$ & $ -6.89$ & $ -5.50$ \\
            & Fe & $  -8.56$ & $ -7.13$ & $ -5.79$ & $ -4.40$ \\
\hline
            & He & $ -16.40$ & $-12.29$ & $ >-9.96$ & $ >-7.17$ \\
            &  O & $  -8.13$ & $ -6.70$ & $ -5.36$ & $ -3.96$ \\
            & Mg & $  -8.47$ & $ -7.14$ & $ -5.83$ & $ -4.45$ \\
    20000 K & Si & $  -8.50$ & $ -7.16$ & $ -5.85$ & $ -4.48$ \\
            & Ca & $  -9.72$ & $ -8.36$ & $ -7.05$ & $ -5.66$ \\
            & Fe & $  -8.66$ & $ -7.26$ & $ -5.95$ & $ -4.56$ \\
\hline
            & He & $ -16.15$ & $-12.20$ & $ >-9.46$ & $ >-6.59$ \\
            &  O & $  -8.26$ & $ -6.83$ & $ -5.53$ & $ -4.14$ \\
            & Mg & $  -8.57$ & $ -7.27$ & $ -6.01$ & $ -4.62$ \\
    25000 K & Si & $  -8.61$ & $ -7.30$ & $ -6.03$ & $ -4.65$ \\
            & Ca & $  -9.84$ & $ -8.50$ & $ -7.23$ & $ -5.83$ \\
            & Fe & $  -8.76$ & $ -7.39$ & $ -6.13$ & $ -4.73$ \\
\hline
\end{tabular}
\end{table}

\begin{table}   
\caption{Same as Table \ref{table:2} for models with $M_\mathrm{H}=10^{-8} M_\odot$}
\label{table:4}
\centering
\begin{tabular}{cccccc}
\hline
            &     & 10$^6$ g s$^{-1}$ & 10$^8$ g s$^{-1}$ & 10$^{10}$ g s$^{-1}$  & 10$^{12}$ g s$^{-1}$\\
\hline
            & He & $  -8.65$ & $ -6.63$ & $ -4.81$ & $ >-2.70$ \\
            &  O & $  -8.62$ & $ -6.72$ & $ -5.45$ & $ -3.99$ \\
            & Mg & $  -8.81$ & $ -7.03$ & $ -5.89$ & $ -4.44$ \\
     8000 K & Si & $  -8.92$ & $ -7.07$ & $ -5.92$ & $ -4.48$ \\
            & Ca & $ -10.12$ & $ -8.30$ & $ -7.15$ & $ -5.68$ \\
            & Fe & $  -9.10$ & $ -7.27$ & $ -6.04$ & $ -4.59$ \\
\hline
            & He & $ -10.24$ & $ -6.81$ & $ -4.78$ & $ >-3.96$ \\
            &  O & $  -8.35$ & $ -6.60$ & $ -5.26$ & $ -3.92$ \\
            & Mg & $  -8.62$ & $ -6.89$ & $ -5.64$ & $ -4.37$ \\
    10000 K & Si & $  -8.65$ & $ -6.93$ & $ -5.68$ & $ -4.40$ \\
            & Ca & $  -9.91$ & $ -8.19$ & $ -6.91$ & $ -5.59$ \\
            & Fe & $  -8.85$ & $ -7.12$ & $ -5.79$ & $ -4.49$ \\
\hline
            & He & $ -10.27$ & $ -6.77$ & $ -4.77$ & $ >-2.60$ \\
            &  O & $  -8.32$ & $ -6.59$ & $ -5.25$ & $ -3.78$ \\
            & Mg & $  -8.61$ & $ -6.89$ & $ -5.62$ & $ -4.23$ \\
    10500 K & Si & $  -8.60$ & $ -6.90$ & $ -5.64$ & $ -4.26$ \\
            & Ca & $  -9.90$ & $ -8.19$ & $ -6.89$ & $ -5.46$ \\
            & Fe & $  -8.83$ & $ -7.11$ & $ -5.77$ & $ -4.35$ \\
\hline
            & He & $ -10.54$ & $ -6.82$ & $ -4.80$ & $ >-2.62$ \\
            &  O & $  -8.32$ & $ -6.61$ & $ -5.20$ & $ -3.72$ \\
            & Mg & $  -8.65$ & $ -6.94$ & $ -5.61$ & $ -4.19$ \\
    11000 K & Si & $  -8.59$ & $ -6.90$ & $ -5.59$ & $ -4.20$ \\
            & Ca & $  -9.94$ & $ -8.25$ & $ -6.90$ & $ -5.42$ \\
            & Fe & $  -8.86$ & $ -7.16$ & $ -5.77$ & $ -4.31$ \\
\hline
            & He & $ -10.08$ & $ -6.51$ & $ -4.71$ & $ >-2.70$ \\
            &  O & $  -8.07$ & $ -6.44$ & $ -5.11$ & $ -3.73$ \\
            & Mg & $  -8.41$ & $ -6.81$ & $ -5.55$ & $ -4.18$ \\
    11500 K & Si & $  -8.37$ & $ -6.82$ & $ -5.57$ & $ -4.20$ \\
            & Ca & $  -9.66$ & $ -8.06$ & $ -6.81$ & $ -5.43$ \\
            & Fe & $  -8.58$ & $ -6.99$ & $ -5.70$ & $ -4.32$ \\
\hline
            & He & $  -9.73$ & $ -6.34$ & $ -4.64$ & $ >-3.60$ \\
            &  O & $  -7.92$ & $ -6.33$ & $ -5.00$ & $ -3.64$ \\
            & Mg & $  -8.25$ & $ -6.72$ & $ -5.47$ & $ -4.12$ \\
    12000 K & Si & $  -8.22$ & $ -6.74$ & $ -5.49$ & $ -4.15$ \\
            & Ca & $  -9.49$ & $ -7.94$ & $ -6.70$ & $ -5.33$ \\
            & Fe & $  -8.42$ & $ -6.87$ & $ -5.59$ & $ -4.23$ \\
\hline
            & He & $  -9.11$ & $ -6.09$ & $ -4.61$ & $ >-3.74$ \\
            &  O & $  -8.01$ & $ -6.44$ & $ -5.18$ & $ -4.75$ \\
            & Mg & $  -8.33$ & $ -6.86$ & $ -5.67$ & $ -4.54$ \\
    16000 K & Si & $  -8.35$ & $ -6.88$ & $ -5.69$ & $ -4.44$ \\
            & Ca & $  -9.59$ & $ -8.08$ & $ -6.90$ & $ -6.34$ \\
            & Fe & $  -8.54$ & $ -6.98$ & $ -5.79$ & $ -5.01$ \\
\hline
            & He & $  -8.73$ & $ -5.91$ & $ >-4.62$ & $ >-4.04$ \\
            &  O & $  -8.11$ & $ -6.55$ & $ -5.36$ & $ -4.92$ \\
            & Mg & $  -8.39$ & $ -7.02$ & $ -5.86$ & $ -4.67$ \\
    20000 K & Si & $  -8.42$ & $ -7.04$ & $ -5.88$ & $ -4.57$ \\
            & Ca & $  -9.66$ & $ -8.21$ & $ -7.07$ & $ -6.51$ \\
            & Fe & $  -8.62$ & $ -7.11$ & $ -5.97$ & $ -5.15$ \\
\hline
            & He & $  -8.40$ & $ -5.80$ & $ >-4.70$ & $ >-3.91$ \\
            &  O & $  -8.22$ & $ -6.68$ & $ -5.55$ & $ -4.23$ \\
            & Mg & $  -8.47$ & $ -7.22$ & $ -6.07$ & $ -4.72$ \\
    25000 K & Si & $  -8.51$ & $ -7.24$ & $ -6.09$ & $ -4.75$ \\
            & Ca & $  -9.76$ & $ -8.38$ & $ -7.28$ & $ -5.93$ \\
            & Fe & $  -8.70$ & $ -7.28$ & $ -6.18$ & $ -4.82$ \\
\hline
\end{tabular}
\end{table}

\begin{table}   
\caption{Same as Table \ref{table:2} for models with $M_\mathrm{H}=10^{-10} M_\odot$}
\label{table:5}
\centering
\begin{tabular}{cccccc}
\hline
            &     & 10$^6$ g s$^{-1}$ & 10$^8$ g s$^{-1}$ & 10$^{10}$ g s$^{-1}$ \\                       
\hline
            & He & $  -7.59$ & $ -6.35$ & $ -4.49$ \\
            &  O & $  -8.31$ & $ -6.47$ & $ -4.87$ \\
            & Mg & $  -8.60$ & $ -6.68$ & $ -4.94$ \\
    10000 K & Si & $  -8.55$ & $ -6.71$ & $ -4.97$ \\
            & Ca & $  -9.88$ & $ -8.03$ & $ -6.37$ \\
            & Fe & $  -8.80$ & $ -6.94$ & $ -5.23$ \\
\hline
            & He & $  -8.31$ & $ -6.30$ & $ -4.49$ \\
            &  O & $  -8.28$ & $ -6.48$ & $ -4.83$ \\
            & Mg & $  -8.58$ & $ -6.70$ & $ -4.95$ \\
    10500 K & Si & $  -8.50$ & $ -6.66$ & $ -4.86$ \\
            & Ca & $  -9.86$ & $ -8.05$ & $ -6.38$ \\
            & Fe & $  -8.78$ & $ -6.96$ & $ -5.24$ \\
\hline
            & He & $  -8.76$ & $ -6.54$ & $ -4.50$ \\
            &  O & $  -8.55$ & $ -6.80$ & $ -4.86$ \\
            & Mg & $  -8.86$ & $ -7.02$ & $ -5.17$ \\
    11000 K & Si & $  -8.70$ & $ -6.88$ & $ -4.98$ \\
            & Ca & $ -10.16$ & $ -8.37$ & $ -6.48$ \\
            & Fe & $  -9.07$ & $ -7.28$ & $ -5.36$ \\
\hline
            & He & $  -8.21$ & $ -5.98$ & $ -4.15$ \\
            &  O & $  -8.03$ & $ -6.24$ & $ -4.50$ \\
            & Mg & $  -8.34$ & $ -6.48$ & $ -4.69$ \\
    11500 K & Si & $  -8.24$ & $ -6.48$ & $ -4.76$ \\
            & Ca & $  -9.59$ & $ -7.75$ & $ -6.06$ \\
            & Fe & $  -8.51$ & $ -6.71$ & $ -5.02$ \\
\hline
            & He & $  -9.68$ & $ -6.96$ & $ -3.74$ \\
            &  O & $  -9.76$ & $ -7.87$ & $ -4.27$ \\
            & Mg & $  -9.21$ & $ -7.17$ & $ -4.65$ \\
    12000 K & Si & $  -9.16$ & $ -7.17$ & $ -4.68$ \\
            & Ca & $ -10.95$ & $ -8.98$ & $ -5.90$ \\
            & Fe & $  -9.75$ & $ -7.82$ & $ -4.81$ \\
\hline
            & He & $  -9.55$ & $ -7.04$ & $ -3.45$ \\
            &  O & $  -9.73$ & $ -7.59$ & $ -4.26$ \\
            & Mg & $  -9.40$ & $ -7.22$ & $ -4.65$ \\
    16000 K & Si & $  -9.16$ & $ -7.04$ & $ -4.67$ \\
            & Ca & $ -11.33$ & $ -9.15$ & $ -5.89$ \\
            & Fe & $ -10.19$ & $ -8.02$ & $ -4.82$ \\
\hline
            & He & $  -9.56$ & $ -6.93$ & $ -3.38$ \\
            &  O & $  -9.75$ & $ -7.60$ & $ -4.29$ \\
            & Mg & $  -9.40$ & $ -7.22$ & $ -4.73$ \\
    20000 K & Si & $  -9.12$ & $ -6.99$ & $ -4.70$ \\
            & Ca & $ -11.34$ & $ -9.16$ & $ -5.93$ \\
            & Fe & $ -10.20$ & $ -8.03$ & $ -4.85$ \\
\hline
            & He & $  -9.35$ & $ -5.68$ & $ -3.38$ \\
            &  O & $  -9.76$ & $ -5.94$ & $ -4.31$ \\
            & Mg & $  -9.41$ & $ -6.14$ & $ -4.74$ \\
    25000 K & Si & $  -9.03$ & $ -6.22$ & $ -4.73$ \\
            & Ca & $ -11.35$ & $ -7.44$ & $ -5.98$ \\
            & Fe & $ -10.20$ & $ -6.46$ & $ -4.82$ \\
\hline
\end{tabular}
\end{table}

\section{Summary and Discussion}
\label{discussion}
We have presented a series of numerical simulations concerning the 
accretion of material produced by the disintegration of small rocky 
bodies onto DA white dwarfs. These simulations consider the effect 
of the double-diffusive instability, referred to as fingering convection. 
This instability is induced by the inverse $\mu$-gradient resulting from 
the accretion of heavy material on the white dwarf outer layers.
Our simulations are aimed at providing realistic estimates of the 
accretion rates, deduced from the observed heavy elements abundances 
in WD atmospheres, for a large range of effective temperatures, 
hydrogen-mass fractions and accretion rates. The results are presented 
in various graphs and tables, in such a way that the accretion rate may 
easily be deduced from the values of the heavy elements abundances.
When fingering convection is properly considered, the resulting accretion 
rates may be up to several orders of magnitude larger than those estimated 
when ignoring its effect. For given values of the accretion rate and 
effective temperature, the accumulation of heavy elements in the WD 
atmospheres increases for decreasing hydrogen mass-fraction, since the 
fingering convection zone becomes thinner when the H/He transition zone 
is closer to the surface. In the cases of thin hydrogen mass-fraction and 
high accretion rates, the fingering convection may dredge-up some fraction 
of He from the H/He transition zone (see Tables 4 and 5). Such an effect 
produces DABZ type white dwarfs.     
We discuss below the various assumptions that have been adopted in our simulations.

The chemical composition of the accreted material is supposed to be 
similar to the Earth bulk composition. This is the case for most 
observed polluted DA white dwarfs \citep{2019MNRAS.490..202S}. 
There are also evidences of other white dwarfs polluted by material 
with a variety of chemical composition \citep{2012MNRAS.424..333G,2015MNRAS.451.3237W, 
2017ApJ...834....1M} including water-rich and hydrated planetesimals 
\citep{2013Sci...342..218F,2015MNRAS.450.2083R,2020MNRAS.499..171H} and 
volatile-rich planetesimals \citep{2017ApJ...836L...7X}. Our simulations 
are not representative of such cases. 

The accreted material is supposed to be mixed through the convection 
zone and through the fingering convection zone. \cite{2021MNRAS.503.1646C} 
find from their 3D radiation-hydrodynamics simulations that DA white dwarfs 
with effective temperature larger than 13000 K are unable to spread the 
accreted material horizontally on a time scale shorter than the diffusion 
time scale. However these diffusion time scales, estimated at the bottom 
of the convection zone, do not take into account the additional fingering 
convection zone. By considering fingering convection, diffusion happens 
deeper in the star and in a time scale which might be significantly longer. 
The absence of surface abundance variations in polluted DA white dwarfs 
\citep{2008ApJ...677L..43D,2009ApJ...693..697R,2019MNRAS.483.2941W} 
can be explained by 1) horizontal mixing being more efficient 
than predicted, 2) material being accreted in a generally homogeneous 
surface distribution, or 3) observations not being sensitive enough 
to detect variations in abundance. Thus our assumption of 
homogeneously-mixed pollutants in turbulent zones is consistent with observations.

The mixing induced by the fingering convection is supposed to 
be a continuous process. The validity of this assumption has been 
called into question \citep{2015ASPC..493..129K}. However, 
\cite{2015ASPC..493..121B} showed that in the case of a DA white 
dwarf with a hydrogen mass fraction of $M_\mathrm{H}=10^{-4} M_\odot$ 
and $T_\mathrm{eff}=11000$ K, the accretion of a C/O mixture at a rate 
of 9$\times$ $10^9$ g s$^{-1}$ induces the fingering convection. 
\cite{2018ApJ...859L..19B,2019ApJ...872...96B} have considered 
the case of accretion of bulk-Earth composition material and a 
range of accretion rates from $10^5$ g s$^{-1}$ to $10^{12}$ g s$^{-1}$. 
They find that fingering convection develops efficiently for accretion 
rates above approximately $10^6$ g s$^{-1}$ (see \cite{2019ApJ...872...96B} 
their figure 5). Multidimensional simulations accounting for convection 
and fingering mixing are clearly needed to probe how these two physical processes interact. 
The agreement between the results of \cite{2015ASPC..493..121B} and 
of \cite{2018ApJ...859L..19B,2019ApJ...872...96B} and ours gives 
confidence in the way fingering convection is introduced in our simulations.

We hope that our results, with the figures and tables, will be useful to provide 
realistic estimates of the accretion rates, including the effect of 
fingering convection, to be used for further studies on evolved planetary systems.

\begin{acknowledgements}
We would like to thank an anonymous referee for the helpful comments, 
and constructive remarks on this manuscript. This work was supported by 
PICT-2017-0884 from ANPCyT, PIP 112-200801-00940 grant from CONICET, 
and grant G149 from University of La Plata. This research has made use 
of NASA Astrophysics Data System. F.C.W., G.V. and S.V. acknowledge 
financial support from "Programme National de Physique Stellaire (PNPS)" 
of CNRS/INSU, France. 
\end{acknowledgements}  
  
\bibliographystyle{aa} 
\bibliography{paper} 

\end{document}